\documentclass[12pt,thmsa]{article}
\usepackage{sw20lart}

\input{tcilatex}
\begin{document}

\title{Comment on Griffiths about realism, locality and Bell experiments}
\author{Emilio Santos \\
Departamento de F\'{i}sica. Universidad de Cantabria. Santander. Spain}
\maketitle

\begin{abstract}
I argue that quantum mechanics is a realistic theory, but it violates either
strong locality (no superluminal influences) or strict causality (diiferent
effects cannot follow from the same cause).
\end{abstract}

In a recent `comment' Griffiths\cite{Gr} has criticized the statement by
Wiseman\cite{Wis} that the loophole-free violation of a Bell inequality is
the death for local realism. Indeed Griffiths claims that ``quantum theory
itself is both local and realistic when properly interpreted using a quantum
Hilbert space rather than the classical hidden variables''. I believe that
there are people supporting Wiseman and people supporting Griffiths. Indeed
it is an old debate, renewed by the new experiments. The purpose of this
note is to make a short contribution to the debate.

\textbf{Realism}

Experiments in physics usually consist of getting \textit{information} about
physical \textit{systems} in specific \textit{states}. A particular state is
obtained after an appropriate preparation and the information is usually got
via measurments. To be general enough I will consider that the result of the
measurement is statistical, so that the relevant result obtained from the
experiment is a probability $P$. I propose defining \textit{realism} as the
hypothesis that physics makes statements about an external reality
independent of the observers. Thus measurements provide information about
something existing even if nothing is measured. As a consequence a necessary
condition for realism is that the probability of the result, $a$, when the
observable $A$ is measured depends on the observable measured and the state
of the system, $\lambda $, which may be written 
\begin{equation}
P_{\lambda }\left( A=a\right) =P\left( a;A,\lambda \right) .  \label{0}
\end{equation}
This is valid in both classical and quantum physics. The classical case is
well known. But eq.$\left( \ref{0}\right) $ is also valid for a quantum
system provided that $\lambda $ represents a vector $\psi $ in the Hilbert
space, or more generally a density operator $\hat{\rho}$. In fact quantum
mechanics predicts the following probability density, $\rho \left( a\right) $
for the (continuous) variable $a$%
\begin{equation}
\rho \left( a\right) =\frac{1}{2\pi }\int d\zeta \exp \left( -i\zeta
a\right) \left\langle \psi \left| \exp \left( i\zeta \hat{A}\right) \right|
\psi \right\rangle ,  \label{1}
\end{equation}
$\hat{A}$ being the quantum operator associated to the observable $A.$

I have written eq.$\left( \ref{0}\right) $ for a discrete variable $a$
because it is easier to understand, but eq.$\left( \ref{1}\right) $ for a
continuous variable because it is a standard quantum formula. For the sake
of simplicity in this note I will consider observables with values $\left\{
0,1\right\} $ where the probability agrees with the expectation, that is

\begin{equation}
P\left( A=1\right) =\left\langle A\right\rangle =\left\langle \psi \left| 
\hat{A}\right| \psi \right\rangle .  \label{2}
\end{equation}

I conclude that \textit{both classical and quantum physics are realistic
theories} in the sense above defined.

\textbf{Causality vs. completeness}

Since the foundational days a debate took place concerning completeness vs.
incompleteness of quantum mechanics, Bohr and Einstein respectively being
the most conspicuous proponents. Here I will comment on a related, and
deeper, question namely whether there is strict causality in nature. The
history of science has always being the search for causal connections
between events and therefore it is hard to believe that strict causality
does not hold true. However there are empirical facts that have led many
people to propose that strict causality is not valid at the microscopic
level. If strict causality does not hold, then necessarily quantum mechanics
should be incomplete.

It is a fact that the results of several runs of a measurement have a
dispersion even if the preparation of the state and the measuring set-up are
identical, as far as we can control. In classical physics there is
dispersion too, but it is usually small. In contrast in quantum physics the
dispersion may be quite large. There are two possible interpretations of
this fact. If we support strict causality in the natural world then the
dispersion of the measurement results should be a consequence of incomplete
information, due to lack of control of all possible variables in the
preparation or the measurement. This is the usual assumption in classical
physics.\ However in the quantum case many people believes that the
dispersion is a consequence of the absence of strict causality in nature,
that is the existence of a kind of fundamental randomness such that
different effects may follow from precisely the same cause.

If we assume strict causality we should write the probability eq.$\left( \ref
{1}\right) $ in the form

\begin{equation}
P=\sum_{\lambda }w_{\lambda }P_{\lambda }(A=a),w_{\lambda }\geq
0,\sum_{\lambda }w_{\lambda }=1,  \label{a2}
\end{equation}
where $\left\{ \lambda \right\} $ is a set of `hidden' states and $\left\{
w_{\lambda }\right\} $ the associated weights. That is we should suppose
that \textit{quantum mechanics is incomplete}\cite{EPR} and might be
completed with subquantum (real, ontic) states, $\left\{ \lambda \right\} .$
(A more popular name for $\lambda $ is hidden variable). It is the case that
the quantum formalism is compatible with eq.$\left( \ref{a2}\right) $ in
simple instances. In fact we may introduce in eq.$\left( \ref{2}\right) $ a
resolution of the identity 
\begin{equation}
1=\sum_{\lambda }\mid \lambda \rangle \langle \lambda \mid ,  \label{a3}
\end{equation}
where $\left\{ \mid \lambda \rangle \right\} $ is a complete orthonormal set
of eigenstates of $\hat{A}$ with eigenvalues $\left\{ \lambda \right\} $,
leading to 
\begin{equation}
P\left( A=1\right) =\sum_{\lambda }\delta _{\lambda 1}\left| \langle \lambda
\mid \psi \rangle \right| ^{2},  \label{a4}
\end{equation}
which agrees with eq.$\left( \ref{a2}\right) $ , the quantities $\left|
\langle \lambda \mid \psi \rangle \right| ^{2}$ playing the role of $%
w_{\lambda }$ and the Kronecker delta $\delta _{\lambda 1}$ the role of $%
P_{\lambda }(A=1).$

I have illustrated the argument with a simple example, but the
generalization to several arbitrary \textit{commuting} observables is
straightforward. For instance if we have two observables $A$ and $B$ whose
associated operator commute, the probability that both have the value $1$
equals the correlation, that is 
\begin{equation}
P\left( A=B=1\right) =\left\langle \psi \left| \hat{A}\hat{B}\right| \psi
\right\rangle .  \label{a5}
\end{equation}
The existence of a complete set of orthonormal eigenstates $\left\{ \mid
\lambda \rangle \right\} $ is guaranteed by the commutativity of the
operators and similar steps as those above lead to 
\begin{eqnarray}
\left\langle \psi \left| \hat{A}\hat{B}\right| \psi \right\rangle
&=&\sum_{\lambda }\left\langle \psi \left| \hat{A}\mid \lambda \rangle
\langle \lambda \mid \hat{B}\right| \psi \right\rangle  \nonumber \\
&=&\sum_{\lambda }\left| \left\langle \psi \mid \lambda \right\rangle
\right| ^{2}\left\langle \lambda \left| \hat{A}\mid \lambda \rangle \langle
\lambda \mid \hat{B}\right| \lambda \right\rangle  \nonumber \\
&=&\sum_{\lambda }\left| \left\langle \psi \mid \lambda \right\rangle
\right| ^{2}\delta _{\lambda 1},  \label{a6}
\end{eqnarray}
where I have taken into account that $\mid \lambda \rangle $ is an
eigenvector of the projector $\hat{A}$ with eigenvalue either $1$ or $0$,
and similar for $\hat{B}$.

In summary if we support strict causality we should assume that quantum
mechanics is incomplete and eqs.$\left( \ref{a2}\right) $ to $\left( \ref{a6}%
\right) $ may be interpreted as an explanation for the dispersion of the
measurement results. In contrast if we support fundamental randomness, no
explanation is needed and those equations may be seen as just formal
developments devoid of any physical interpretation.

Strict causality is closely related to \textit{determinism}, but some
authors make a distinction. In fact due to some unavoidable noise, usually
attributed to vacuum fluctuations\cite{FOS}, we might have strict causality
but practical randomness, i. e. practical lack of determinism. But I shall
not discuss this question anymore here.

\textbf{Locality}

The question of locality involves measurements made, by Alice and Bob, in
distant places (or more strictly in space-like separated regions in the
sense of relativity theory). Thus we consider a system consisting of two
separated subsystems in the global state $\psi $ and two measurements, one
on each subsystem, corresponding to the observables $A$ and $B$. We will
assume that these observables are defined for Alice's and Bob's subsystems
respectively. Both observables are jointly measurable if the associated
operators commute, which is always true if the measurements are space-like
separated. In fact quantum field theory postulates that the field operators
commute in that case, and we should assume that the operators $\hat{A}$ and $%
\hat{B}$ are functions of the field operators. Eqs.$\left( \ref{a5}\right) $
and $\left( \ref{a6}\right) $ are still valid and the only new feature is
that $A$ and $B$ are uncorrelated if the state vector $\psi $ factorizes,
one term being attached to Alice's subsystem and the other one to Bob's. In
contrast, quantum correlations amongst observables of separated systems
appear if $\psi $ does not factorize, what gives rise the a socalled \textit{%
entangled} state.

A sharp difference between classical and quantum correlations appears when
there are noncommuting operators. Let us consider three observables, $A,B,C,$
whose associated quantum operators have the following commutation properties 
\begin{equation}
\left[ \hat{A},\hat{B}\right] =0,\left[ \hat{A},\hat{C}\right] =0,\left[ 
\hat{B},\hat{C}\right] \neq 0.  \label{9}
\end{equation}
We assume that $A$ is a property of Alice's subsystem whilst both $B$ and $C$
are properties of Bob's subsystem. For any correlation between $A$ and $B$
we may arrive at a construction like eq.$\left( \ref{a6}\right) .$ Similarly
any correlation between $A$ and $C$ may be written in a similar form, namely 
\begin{equation}
\left\langle \psi \left| \hat{A}\hat{C}\right| \psi \right\rangle =\sum_{\mu
}\left| \left\langle \psi \mid \mu \right\rangle \right| ^{2}\left\langle
\mu \left| \hat{A}\mid \mu \rangle \langle \mu \mid \hat{C}\right| \mu
\right\rangle ,  \label{10}
\end{equation}
where $\left\{ \mid \mu \rangle \right\} $ is a complete orthonormal set of
simultaneous eigenstates of $\hat{A}$ and $\hat{C}$.

However a problem arises, namely the states, $\mid \mu \rangle ,$ of the
system involved in eq.$\left( \ref{10}\right) $ are different from those, $%
\mid \lambda \rangle $ , involved in eq.$\left( \ref{a6}\right) .$ No simple
physical explanation may be given to this formal feature. A possible
interpretation might be that both the quantum states $\mid \lambda \rangle $
and $\mid \mu \rangle $ are actually mixtures of some set of `subquantum'
states, say $\nu ,$ such that all quantum states may be written in terms of
them. That is 
\begin{equation}
\lambda =\sum_{\nu }p\left( \nu ;\lambda \right) \nu ,\text{ }\mu =\sum_{\nu
}p\left( \nu ;\mu \right) \nu ,  \label{11}
\end{equation}
this simbolic expression meaning that $\lambda $ (the quantum state
represented by $\mid \lambda \rangle $ in the Hilbert space formalism) is a
mixture of the subquantum states $\nu $ with weights $p\left( \nu ;\lambda
\right) ,$ and similar for $\mu .$ If this is the case, taking eq.$\left( 
\ref{11}\right) $ into account both correlations $\left\langle \psi \left| 
\hat{A}\hat{B}\right| \psi \right\rangle $ and $\left\langle \psi \left| 
\hat{A}\hat{C}\right| \psi \right\rangle $ could be written in the
classical-like form eq.$\left( \ref{2}\right) .$ However this is not
possible in general: as a consequence of Bell's theorem\cite{Bell} no set $%
\left\{ \nu \right\} $ exist allowing to express both correlations in terms
of states of this set. In fact, any value of $\nu $ would attach a definite
value to every one of the three observables $A,B$ and $C$, thus givin rise
to a joint probability distribution of a set of observables (with values $%
\left\{ 0,1\right\} ).$ But the existence of such a distribution implies the
fulfillement of all Bell inequalities\cite{Santos}, which does not hold true
in some cases. Thus there seems to be some \textit{nonlocal (superluminal)
influence} between Alice's and Bob's subsystems. We migh say that a \textit{%
strong }form of\textit{\ locality} (`superluminal influences' do not exist)
is violated. However a \textit{weak }form of\textit{\ locality} is
compatible with quantum mechanics, which is usually named `no signalling'%
\cite{Bed}. It forbids only superluminal signals, but not superluminal
influences. (I use weak and strong in the sense that strong$\Rightarrow $%
weak).

\textbf{Conclusions}

1. Quantum mechanics is a realistic theory.

2. Quantum mechanics either violates \textit{strict causality} or it
violates at least a \textit{strong} form of \textit{locality}. The
alternative is dramatic because both terms of the dilemma are cherised
principles of classical physics. Indeed Einstein supported both.

If we pass from the quantum \textit{theory} to the \textit{empirical facts},
the same alternative appears provided a loophole-free violation of a Bell
inequality is produced, which seems to be the case according to recently
reported experimental results.

\end{document}